\documentclass[unnumsec,webpdf,contemporary,large,namedate]{oup-authoring-template}% uncomment this line for author year citations and comment the above
\onecolumn % for one column layouts

\usepackage{amsmath,amsthm,amssymb,amsfonts}
\usepackage{mathtools}
\usepackage{graphicx}
\usepackage{verbatim}
\usepackage{listings}
\usepackage{algorithm}
\usepackage{csquotes}
\usepackage[english]{babel}
\usepackage[hidelinks]{hyperref}
\usepackage{adjustbox}
\usepackage{multirow}
\usepackage{pdflscape}
\usepackage{enumerate}
\usepackage{url} 
\usepackage{tkz-graph} 
\usetikzlibrary{shapes.geometric, arrows.meta, arrows}
\tikzstyle{VertexStyle} = [shape            = ellipse,
                            minimum width    = 6ex,%
                            draw]

\tikzstyle{EdgeStyle}   = [->,>=stealth']

\newcommand{\iid}{\overset{iid}{\sim}}

\graphicspath{{Figures/}}

\theoremstyle{thmstyleone}%
\newtheorem{theorem}{Theorem}%  meant for continuous numbers
\theoremstyle{thmstyletwo}%
\theoremstyle{thmstylethree}%

\newtheorem{assumption}{Assumption}

\begin{document}

\vspace{10pc}
\begin{center}
	\large
	Homophily-adjusted social influence estimation\\
	\normalsize
	
	\vspace{5pc}
	Hanh T.D. Pham \\
	Department of Biostatistics, University of Iowa\\
	
	\vspace{2pc}
	Daniel K. Sewell \\
	Department of Biostatistics, University of Iowa\\
	145 N. Riverside Dr. \\
	Iowa City, IA 52242 USA \\
	\url{daniel-sewell@uiowa.edu}
	
\end{center}

%\newpage

\journaltitle{Journal Title Here}
\DOI{DOI HERE}
\copyrightyear{2022}
\pubyear{2019}
\access{Advance Access Publication Date: Day Month Year}
\appnotes{Paper}

\firstpage{1}

%\subtitle{Subject Section}

\title[Homophily-adjusted social influence estimation]{Homophily-adjusted social influence estimation}

\author[1]{Hanh T.D. Pham}
\author[1,$\ast$]{Daniel K. Sewell}

\authormark{Pham et al.}

\address[1]{\orgdiv{Department of Biostatistics}, \orgname{Organization}, \orgaddress{\street{145 N. Riverside Dr., Iowa City}, \postcode{52242}, \state{IA}, \country{USA}}}

\corresp[$\ast$]{Corresponding author. \href{email:daniel-sewell@uiowa.edu}{daniel-sewell@uiowa.edu}}

\received{Date}{0}{Year}
\revised{Date}{0}{Year}
\accepted{Date}{0}{Year}

%\editor{Associate Editor: Name}

%\abstract{
%\textbf{Motivation:} .\\
%\textbf{Results:} .\\
%\textbf{Availability:} .\\
%\textbf{Contact:} \href{name@email.com}{name@email.com}\\
%\textbf{Supplementary information:} Supplementary data are available at \textit{Journal Name}
%online.}

\abstract{
Homophily and social influence are two key concepts of social network analysis. Distinguishing between these phenomena is difficult, and approaches to disambiguate the two have been primarily limited to longitudinal data analyses. In this study, we provide sufficient conditions for valid estimation of social influence through cross-sectional data, leading to a novel homophily-adjusted social influence model which addresses the backdoor pathway of latent homophilic features. The oft-used network autocorrelation model (NAM) is the special case of our proposed model with no latent homophily, suggesting that the NAM is only valid when all homophilic attributes are observed. We conducted an extensive simulation study to evaluate the performance of our proposed homophily-adjusted model, comparing its results with those from the conventional NAM. Our findings shed light on the nuanced dynamics of social networks, presenting a valuable tool for researchers seeking to estimate the effects of social influence while accounting for homophily. Code to implement our approach is available at \url{https://github.com/hanhtdpham/hanam}.
}

% \boxedtext{
% \begin{itemize}
% \item Key boxed text here.
% \item Key boxed text here.
% \item Key boxed text here.
% \end{itemize}}

\maketitle

\section{Introduction}
Social influence and homophily are two fundamental concepts in social network analysis that shed light on the dynamics and structure of connections. Social influence refers to the phenomenon where individuals in a network are affected by the actions, opinions, and behaviors of their connections. This influence can manifest in various forms, such as adopting similar beliefs, attitudes, or behaviors due to interaction and communication within the network. In network literature, this phenomenon is also known as ``social contagion,'' ``diffusion,'' and ``peer influence.'' On the other hand, homophily refers to the tendency of individuals forming bonds with others who share similar characteristics, interests, or traits. Separating social influence from homophily is challenging. The two phenomenon are generally confounded with each other in observational data \citep{shalizi2011}. For example, if we observe two people being friends and they both smoke, it is difficult to determine whether the friendship influenced the behavior or vice versa. 

Nevertheless, some researchers have proposed approaches to account for homophily when estimating social influence within longitudinal data contexts. \cite{steglich2010} adopted an agent-based approach to modeling the co-evolution of homophily and influence, supported by the award winning software Siena \citep{rsiena}. \cite{xu2018} considered the failure to account for homophily in estimating influence as an omitted variable bias problem and proposed three methods for adjustment. Under the framework of linear models, \cite{mcfowland2023} used latent location estimates from either a latent community model or a continuous latent space model as a proxy for homophily.

Distinguishing social influence from homophily is even more challenging in cross-sectional data. Current practice to estimate social influence is to use the network autocorrelation model (NAM) \citep{ord1975estimation,doreian1989network}, which despite its long-standing history, fully ignores unobserved homophily. As we will show later, due to lack of information about the sequence of events, strong assumptions are necessary to disentangle the effect of social influence versus homophily. In this paper, we propose two homophily-adjusted network influence models for cross-sectional data. We first derive our models via rigorous processes of establishing longitudinal linear models and then specify the required conditions for the model to converge to a stable limiting distribution.  Under these sufficient conditions, our approach allows valid estimation of social influence without bias induced by the backdoor pathway of homophilic features.  We later provide a computationally efficient method for estimation and inference that is, as will be made clear later, flexible to a variety of network models. 

The remainder of our paper is organized as follows. Section \ref{sec:NAM} provides background on the network autocorrelation model, its justification, and the assumptions implicit in its use.  Section \ref{sec:hanam} presents the details of our model and estimation approach, along with the assumptions required for valid estimation. Section \ref{sec:simstudy} comprises a simulation study to evaluate the performance of our model. Section \ref{sec:application} illustrates an example of applying our method to a real data set to measure the effect of peer influence on physical activity. Finally, Section \ref{sec:discussion} concludes with a discussion of the paper's findings, highlighting the limitations of the proposed method and suggesting potential avenues for future research.

\section{Network Autocorrelation Models}
\label{sec:NAM}
We commence with an introduction to the standard framework used for quantifying social influence from cross-sectional data, known as the network autocorrelation model (NAM), and an in depth look at the NAM's underlying assumptions. The NAM involves three components: the static network, predictor variables, and the outcome variable of interest.  The predictor variables and outcome variable are related through a linear regression model with adjustments to the mean structure, the covariance structure, or both due to the underlying network connecting the subjects. The two most commonly used NAMs are the network effects model and network disturbances model \citep{doreian1980}, and will be described below. 

The notation to be used is as follows. Let $A$ denote the adjacency matrix corresponding to a static network involving $n$ individuals. Historically, $A$ is taken to be row-normalized, and we will proceed similarly throughout this paper. Let $y$ denote the $n\times1$ continuous outcome vector and $X$ denote the $n \times p$ design matrix of observed covariates. Let the subscript $_t$ denote the measurement at time $t$.  

The network effects model is given by
\begin{align}
    y  
    & 
    = X\beta + \rho A y + \epsilon, \nonumber
    \\
    \epsilon 
    & 
    \sim {{\mathcal N}}(0, \sigma^2 I),\label{eq:NAM_effect}
\end{align}
where ${{\mathcal N}}(\mu, \Sigma)$ denotes the multivariate normal distribution with mean $\mu$ and covariance $\Sigma$, and $I$ is the identity matrix. This model assumes that an individual's mean response is a function of not only their observed characteristics but also the responses of their neighbors. Equivalently (and more conveniently for performing estimation and inference), we can express the model through the likelihood given by
\begin{align}
     y |A, X \sim {{\mathcal N}}\Big((I-\rho A)^{-1} X\beta, \sigma^2 (I-\rho A)^{-1}(I-\rho A')^{-1}\Big).
     \label{NAM_effect_dist}
\end{align}

The justification of using the network effects model lies in the equilibrium solution of a discrete-time model describing the evolution of actors' behavior within a fixed network \citep{butts2023}. The discrete-time model is represented by the recursive equation:
\begin{align}
    y_t = \rho A y_{t-1} + z,
    \label{eq:linear_diffusion}
\end{align}
where $z$ represents a feature vector for the individuals. This linear diffusion model converges only if the following assumption is satisfied:
\begin{assumption}
    \label{assump:convergence}
    % \textit{(Convergence)} 
    $\lambda_1(\rho A) < 1$,
\end{assumption}
\noindent where $\lambda_1(\cdot)$ is the largest singular value, i.e., the spectral norm of $\rho A$.

Under Assumption \ref{assump:convergence}, $y_t$ converges to a stable fixed point as $t \to \infty$. The equilibrium solution is given by $y:=\lim_{t\to\infty}y_t = (I - \rho A)^{-1}z$ \citep{butts2023}. The network effects model consists of this solution setting $z := X\beta + \epsilon$.  Implicitly, then, the NAM makes the additional following assumptions:

\begin{assumption}
    \label{assump:stability_network}
    % \textit{(Network stability)} 
    There exists a time point $t^*$ such that there are negligible network dynamics for all $t>t^*$, i.e., $A_t = A$ $\forall t>t^*$.
\end{assumption}
\noindent Without loss of generality, we will denote this time point of stability as $t^*=0$.  
\begin{assumption}
    \label{assump:nam_finite}
    % \textit{(Finitiness)} 
    $\|y_0\|_\infty < \infty$,
\end{assumption}
\noindent where $\|\cdot\|_{\infty}$ is the infinity norm. 
\begin{assumption}
    \label{assump:nam_deterministic}
    % \textit{(Deterministic process)}
    Given the actor features $z$ and the outcome values at time 0, $y_0$, the social influence process is deterministic.
\end{assumption}
\begin{assumption}
    \label{assump:no_unmeasured_covariates_nam}
    % \textit{(Unmeasured confounders)} 
    There are no unmeasured covariates outside of $X$ which affect both the outcome $y$ and the network.
\end{assumption}
\noindent Importantly, if Assumption \ref{assump:no_unmeasured_covariates_nam} is violated, then there exists a backdoor pathway that will confound the estimation, leading to biased estimates of the social influence.

The second commonly used NAM is the network disturbances model and is given by
\begin{align}
    y = X\beta + \nu, \nonumber\\
    \nu = \rho A \nu + \epsilon,\nonumber\\
    \epsilon \sim {{\mathcal N}}(0, \sigma^2 I).\label{NAM_disturbance}
\end{align}
In this model, an individual's deviance from their mean response is a function of their neighbors' deviances. The corresponding likelihood for the disturbances model is given by
\begin{align}
     y | A, X \sim {{\mathcal N}}\Big(X\beta, \sigma^2_\alpha (I-\rho A)^{-1}(I-\rho A')^{-1}\Big).
     \label{NAM_disturbance_dist}
\end{align}
As with the effects model, the network disturbances model is justified through the linear diffusion model applied to $y_t - X\beta$ rather than $y_t$ and letting $z := \epsilon$.

\section{Homophily-adjusted NAMs}
\label{sec:hanam}
We begin by outlining the theoretical assumptions that underpin the development of a homophily-adjusted social influence model tailored for longitudinal data.  This outcome lays the groundwork for our proposed cross-sectional models for which the NAMs are special cases. We address the challenge posed by unobserved homophily and formulate an estimation approach within a Bayesian framework.

\subsection{Model assumptions}
\label{subsec:assumptions}
\subsubsection*{Network stability}

Assumption \ref{assump:stability_network} indicates that the joint dynamics of the outcome and network have stabilized to the point where the rate of change of the network is negligible compared to the change in the outcome.  This is a strong assumption that must be evaluated very carefully in each application. Yet it is necessary in order to make any attempt at inference on social influence from cross-sectional data.  In practice this means that for the network stability assumption to hold, there must be a plausible point in the past where the network has stabilized and the outcome has continued to evolve.

The linear diffusion model is inherently deterministic, and hence it makes sense in that framework to require that all values of the outcome at the initial time point ($y_0$) are finite (Assumption \ref{assump:nam_finite}). In our approach, we assume that the outcome variable of interest is at all time points stochastic, that is, a random variable. At time $t=0$, $y_0$ and the network $A$ presumably arise from, among other things, some co-evolution stochastic process involving homophily and influence. We therefore replace Assumption 
\ref{assump:nam_deterministic} with the following, which is its stochastic equivalent:
\begin{assumption}
    \label{assump:finiteness}
    % \textit{(Finitiness)}
    There exists $M\in\Re^+$ such that
    $\Pr(\|y_0\|_\infty < M) = 1$,
\end{assumption}
\noindent That is, with probability 1 all elements of $y_0$ are finite.

In short, we assume that the co-evolution process of the network and the actor attributes has achieved a stable form of the network, and at the point at which this has occurred the outcome process has not already diverged.

\subsubsection*{Homophily}
Homophily, the phenomenon that individual level features interact to bring about network connections, is the raison d'\^etre of the broad class of latent space models for networks (LSMs).  LSMs were introduced in the landmark paper by \cite{hoff2002} and have since received much attention theoretically \citep[e.g.,]{rastelli2016properties,lubold2023identifying}, computationally \citep[e.g.,][]{salter2013variational,oconnor2020maximum}, and in practical extensions \citep[e.g.,][]{krivitsky2009representing,sewell2016weighted}.  In LSMs, it is assumed that there are both observed individual-level and/or dyadic-level features as well as unobserved individual-level features which lead to edge formation.  We will denote these unobserved individual level features through the $n\times D$ matrix $U$, where $D$ is an unknown number of latent dimensions. 

These latent, as well as observed, individual features are likely to play a role in the co-evolution process of the network and outcome variable, and although we do assume the coevolution process has achieved a stable network, we do not assume that these unobserved features no longer play a role in the outcome variable.  Therefore, we relax Assumption \ref{assump:no_unmeasured_covariates_nam}.

\subsubsection*{Stochasticity}
One implication of Assumption \ref{assump:nam_deterministic} is that if we observe three consecutive time points we would be able to know $\rho$ and $z$ exactly and predict the outcome variable at any subsequent time point\footnote{Since $y_s = \rho A y_{s-1} + z$, we could exactly compute $\rho$ as $\rho = \frac{(y_{s+1} - y_{s})_i}{(Ay_s - Ay_{s-1})_i}$ $\forall i$.  Computing $z$ and $y_t$, $t>s+1$, immediately follows.}.  As this is not plausible, we wish to relax Assumption \ref{assump:nam_deterministic} by introducing sources of stochasticity. 

We assume that there are two forms of stochasticity involved in the underlying process generating the outcome.  First, there is a persistent, individual-specific stochastic component resulting in deviations from an individual's inherent features ($X$ and $U$) and the influence process; we will denote this stable stochastic component as $\alpha$. Its contribution at each time point $t$ to the outcome $y_t$ is scaled by a quantity we will denote as $\sigma_{\alpha,t}$.  Second, there is an individual- and time-specific stochastic component at each time point $t$, $\epsilon_t$.  That is, $\epsilon_t$ is a random injection of noise occurring at each time point. Its contribution at each time point $t$ to the outcome $y_t$ is scaled by a quantity we will denote as $\sigma_{\epsilon,t}$.  

In the cross-sectional model to be described later, we make the following stability assumption:
\begin{assumption}
    \label{assump:stability_deviations}
    \begin{minipage}[t]{\linewidth}
    \begin{enumerate}[a)]
        \item $|\sigma_{\alpha,t}|, |\sigma_{\epsilon,t}|<\infty$ $\forall t$
        \item $\lim_{t\to\infty} \sigma_{\alpha,t} = \sigma_\alpha<\infty$, and $\lim_{t\to\infty} \sigma_{\epsilon,t} = 0.$
    \end{enumerate}
    \end{minipage}
\end{assumption}
Assumption \ref{assump:stability_deviations} implies that an individual may deviate from that which is predicted from their individual features and social influence, but in a stable manner such that the random deviations eventually become stable. 

We feel that Assumption \ref{assump:stability_deviations} is fairly weak, and it is inherently made whenever an outcome is assumed to be reliable. That is, an individual has reached a point where perfect measurements of an individual's outcome variable will not change from one day to the next (beyond what may be predicted from a change in that individual's covariates). 

Note that we have purposefully not addressed measurement error, as we are describing the true outcome variable of interest.  Any measurement error that may be incurred should be addressed through an additional stochastic component specific to the time of measurement, and, more germane to this paper, should not be used in deriving the limiting distribution of the true underlying outcome variable.  That is, our assumptions are not incompatible with an individual's outcome continuing to vary over time, so long as that continuing temporal variaton is due to measurement error, and not temporal variation in the true measurement error-free outcome.

\subsection{Homophily-adjusted NAM}

Based on the above discussion of Section \ref{subsec:assumptions}, we propose the following longitudinal model to describe the impact of both homophilic attributes and social influence on the outcome of each actor in the network over time for $t\geq 1$.
\begin{align}
    y_t & =  U\gamma + X\beta + \rho A y_{t-1} + \sigma_{\alpha,t}\alpha + \sigma_{\epsilon,t}\epsilon_t, &\nonumber\\
    \alpha & \sim {{\mathcal N}}(0,I_N), \nonumber\\
    \epsilon_t &\iid {{\mathcal N}}(0,I_N). \label{eq:hane_long}
\end{align}
Figure \ref{fig:hnamDAG} depicts a directed acyclic graph illustrating the relationships among variables of actor $i$ and $j$ in the proposed model. It is worth noting that this model shares many similarities with the one proposed in the landmark paper by \cite{mcfowland2023}. Both models rely on Assumption \ref{assump:stability_network}, include the latent locations, the neighbors' outcome at the previous time, and an individual time-specific stochastic component, although the two approaches treat the temporal dependencies somewhat differently.

\begin{landscape}
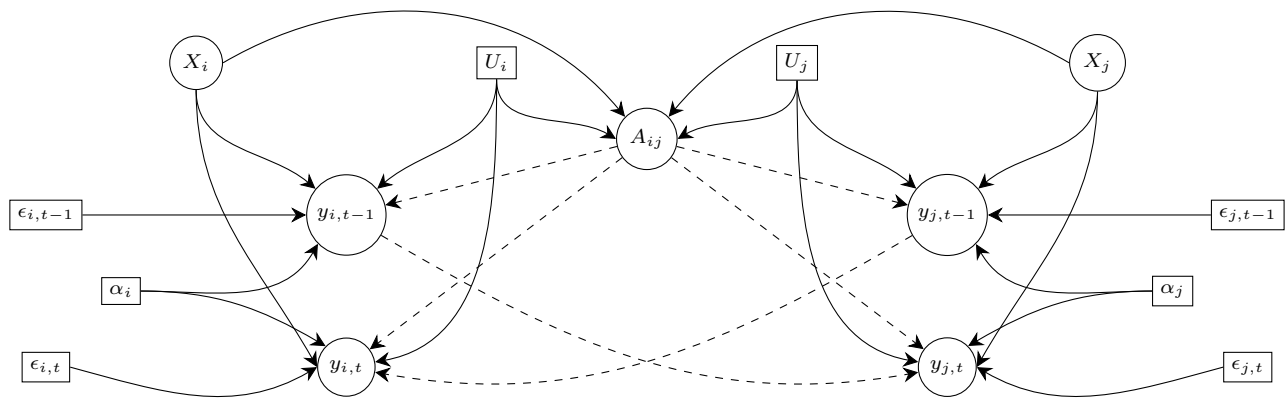
\begin{figure}[hbt]
    \centering
    \begin{tikzpicture}
        % nodes: homophily
        \node[draw, circle]    (Xi)     at (0, 6) {$X_i$};
        \node[draw, rectangle] (Ui)     at (4, 6) {$U_i$};
        \node[draw, circle]    (Xj)     at (12, 6) {$X_j$};
        \node[draw, rectangle] (Uj)     at (8, 6) {$U_j$};

        % nodes: stochastic components
        \node[draw, rectangle]     (ei1) at (-2,4) {$\epsilon_{i,t-1}$};
        \node[draw, rectangle]     (ei2) at (-2,2) {$\epsilon_{i,t}$};
        \node[draw, rectangle]     (ej1) at (14,4) {$\epsilon_{j,t-1}$};
        \node[draw, rectangle]     (ej2) at (14,2) {$\epsilon_{j,t}$};
        \node[draw, rectangle]     (alpha_j) at (13,3) {$\alpha_{j}$};
        \node[draw, rectangle]     (alpha_i) at (-1,3) {$\alpha_{i}$};
        
        % nodes: network
        \node[draw, circle]    (Aij)    at (6, 5) {$A_{ij}$};

        % nodes: outcome
        \node[draw, circle]    (Yit1)   at (2, 4) {$y_{i, t-1}$};
        \node[draw, circle]    (Yit2)   at (2, 2) {$y_{i, t}$};
        \node[draw, circle]    (Yjt1)   at (10, 4) {$y_{j, t-1}$};
        \node[draw, circle]    (Yjt2)   at (10, 2) {$y_{j, t}$};

        % arrows: homophily
        \draw[-{Stealth[length=2mm, width=2mm]}] (Xi.east) to[out = 30, in = 130] (Aij.north west);
        \draw[-{Stealth[length=2mm, width=2mm]}] (Ui.south) to[out = -90, in = 150] (Aij.west);
        \draw[-{Stealth[length=2mm, width=2mm]}] (Xj.west) to[out = 150, in = 50] (Aij.north east);
        \draw[-{Stealth[length=2mm, width=2mm]}] (Uj.south) to[out = -90, in = 30] (Aij.east);
        
        % arrows: influence from A
        \draw[-{Stealth[length=2mm, width=2mm]}, dashed] (Aij) -- (Yit1);
        \draw[-{Stealth[length=2mm, width=2mm]}, dashed] (Aij) -- (Yit2);
        \draw[-{Stealth[length=2mm, width=2mm]}, dashed] (Aij) -- (Yjt1);
        \draw[-{Stealth[length=2mm, width=2mm]}, dashed] (Aij) -- (Yjt2);
        
        % arrows: covariates (i)
        \draw[-{Stealth[length=2mm, width=2mm]}] (Xi) to[out=-90,in=140] (Yit1);
        \draw[-{Stealth[length=2mm, width=2mm]}] (Ui) to[out=-90,in=40] (Yit1);
        \draw[-{Stealth[length=2mm, width=2mm]}] (Xi) to[out=-90,in=120] (Yit2.west);
        \draw[-{Stealth[length=2mm, width=2mm]}] (Ui) to[out=-90,in=10] (Yit2);

        % arrows: covariates (j)
        \draw[-{Stealth[length=2mm, width=2mm]}] (Uj) to[out=-90,in=140] (Yjt1);
        \draw[-{Stealth[length=2mm, width=2mm]}] (Xj) to[out=-90,in=40] (Yjt1);
        \draw[-{Stealth[length=2mm, width=2mm]}] (Xj) to[out=-90,in=60] (Yjt2.east);
        \draw[-{Stealth[length=2mm, width=2mm]}] (Uj) to[out=-90,in=170] (Yjt2);

        % arrows: influence (y to y)
        \draw[-{Stealth[length=2mm, width=2mm]}, dashed] (Yit1) to[out=-30,in=-170] (Yjt2);
        \draw[-{Stealth[length=2mm, width=2mm]}, dashed] (Yjt1) to[out=-150,in=-10] (Yit2);

        % arrows: stochastic components (epsilon)
        \draw[-{Stealth[length=2mm, width=2mm]}] (ei1) to (Yit1);
        \draw[-{Stealth[length=2mm, width=2mm]}] (ei2.east) to[out = -15, in = 210] (Yit2.west);
        \draw[-{Stealth[length=2mm, width=2mm]}] (ej1) to (Yjt1);
        \draw[-{Stealth[length=2mm, width=2mm]}] (ej2.west) to[out = 195, in = -30] (Yjt2.east);

        % arrows: stochastic components (alpha)
        \draw[-{Stealth[length=2mm, width=2mm]}] (alpha_i.east) to[out = 0, in = -120] (Yit1.south west);
        \draw[-{Stealth[length=2mm, width=2mm]}] (alpha_j.west) to[out = 180, in = -60] (Yjt1.south east);
        \draw[-{Stealth[length=2mm, width=2mm]}] (alpha_i.east) to[out = 0, in = 150] (Yit2.north west);
        \draw[-{Stealth[length=2mm, width=2mm]}] (alpha_j.west) to[out = 180, in = 30] (Yjt2.north east);
        
    \end{tikzpicture}
    \caption{A directed acyclic graph showing the relationships among variables in the proposed model. Circles and boxes indicate observed and unobserved variables, respectively. Solid arrows represent the effect of individuals' features. Dashed arrows represent the effect of neighbors.}
    \label{fig:hnamDAG}
\end{figure}
\end{landscape}

Given our primary interest in social influence models for cross-sectional data, we investigate the limiting distribution of the outcome under the model in (\ref{eq:hane_long}) as $t \to \infty$. The following theorem shows that under certain conditions the outcome vector converges in distribution to a multivariate normal distribution. The proof can be found in the Appendix.

\begin{theorem}
\label{theorem:effect}
Under Assumptions \ref{assump:convergence}, \ref{assump:stability_network}, \ref{assump:finiteness}, and \ref{assump:stability_deviations} and Eq. (\ref{eq:hane_long}), we have that
\begin{align}
    y_t\overset{{\mathcal D}}{\to} {{\mathcal N}}\Big((I-\rho A)^{-1} ( U\gamma + X\beta), \sigma^2_\alpha (I-\rho A)^{-1}(I-\rho A')^{-1}\Big).
    \label{eq:hane}
\end{align}
\end{theorem}

The limiting distribution in (\ref{eq:hane}) of the outcome under our proposed model is identical to the network effects model should $\gamma = 0$.  If this is the case, then Theorem \ref{theorem:effect} provides considerably stronger justification for the use of the network effects model, as it is derived from what we feel is a much more reasonable stochastic longitudinal model (Eq. (\ref{eq:hane_long})) than the deterministic linear diffusion model.

If, however, the latent homophilic features are relevant to the outcome of interest (i.e., $\gamma \neq 0$), the latent features $U$ become non-negligible as they provide a backdoor pathway between the outcome and the network. As will be demonstrated in the simulation study, failing to adjust for these homophilic features leads to gross overestimation of the social influence.  

With minor modifications to equations in (\ref{eq:hane_long}), we can derive a homophily-adjusted network disturbances model. Instead of focusing on neighbors' responses, we assume an individual's outcome is a function of a weighted average of their neighbors' deviations at the previous time points. Then the outcome at time $t \geq 1$ is given by
\begin{align}
    y_t & =  U\gamma + X\beta + \rho A (y_{t-1} - U\gamma - X\beta) + \sigma_{\alpha,t}\alpha + \sigma_{\epsilon,t}\epsilon_t.
    \label{eq:hand_long}
\end{align} 
This then yields the following theorem, whose proof is nearly identical to that of Theorem \ref{theorem:effect}.

\begin{theorem}
   Under Assumptions \ref{assump:convergence}, \ref{assump:stability_network}, \ref{assump:finiteness}, and \ref{assump:stability_deviations} and Eq. (\ref{eq:hand_long}) we have that
    \begin{align}
    y_t\overset{{\mathcal D}}{\to} {{\mathcal N}}\Big(U\gamma + X\beta, \sigma^2_\alpha (I-\rho A)^{-1}(I-\rho A')^{-1}\Big).
    \label{eq:disturbance_cond}
\end{align}
\end{theorem}

\subsection{Integrating out latent homophily}
We now address the challenge of appropriately accounting for the unobserved homophily component in the homophily-adjusted network effects and disturbances models. The dimension of the parameter space of Eqs. (\ref{eq:hane}) and (\ref{eq:disturbance_cond}) is very large due to the latent individual features $U$, and it is our goal is to drastically reduce the dimension while remaining model agnostic regarding the joint distribution of $(A,U)$.

Although one could plug in estimates of the latent positions, directly incorporating them into any model requires additional effort to accommodate the bias and additional variances associated with using plug-in estimators. Under a Bayesian framework, we propose integrating out $U$ from the influence models, thereby focusing on the marginal likelihood given by
\begin{align}
    \pi(y|A,X) &= \int \pi(y|U,A,X) \pi(U|A,X)dU.
    \label{eq:marginal_lik}
\end{align}
Note that we have dropped the implicit conditioning on the model parameters for ease of notation. The integration process requires knowledge of $\pi(U|A,X)$, which can be obtained from any of the established LSMs as selected by the practitioner and supported by the data. Rather than specifying a particular process, we accommodate various latent space models, including but not limited to the stochastic block models \citep{nowicki2001estimation} and the latent position models \citep{hoff2002}. This flexibility allows for a broader exploration of potential model frameworks. 

It is improbable that the integral within Eq. (\ref{eq:marginal_lik}) has a closed-form solution. To circumvent this issue, we approximate the posterior $\pi(U|A,X)$ using ideas very similar to a fixed-form variational Bayes approach \citep[see, e.g.,][]{honkela2010approximate}.  To achieve an analytical solution to the integral in (\ref{eq:marginal_lik}), our posterior approximation is constrained to the family of matrix normal distributions, denoted by $\mathcal{MN}_{n\times D} (\Lambda, \Omega, \Psi)$, where $\Lambda$ is the $n\times D$ location matrix, $\Omega$ is the $n\times n$ row covariance matrix, and $\Psi$ is the $D\times D$ column covariance matrix. We assume that posterior samples of $\pi(U|A,X)$ can be obtained, and using these posterior draws we wish to minimize the Kullback-Leibler (KL) divergence between the true posterior distribution and the matrix normal approximation.  That is, letting $q(U|\Lambda,\Omega,\Psi)$ denote the matrix normal approximation of the true posterior $\pi(U|A,X)$, we wish to find
\begin{align}
    KL\big( \pi \| q \big) & = \mbox{constant} - \mathbb{E}_\pi\big(q(U|\Lambda,\Omega,\Psi) | A,X\big). &
    \label{eq:kl}
\end{align}

By evaluating the expectation in Eq. (\ref{eq:kl}) via Monte Carlo using $K$ posterior draws $(U^{(1)}, \ldots, U^{(K)})$ from $\pi(U|A,X)$, minimizing the KL divergence is equivalent to setting the the location matrix, the row and column covariance matrix to the maximum likelihood estimates of the sample.  
Consequently, we obtain the following estimates for $\Lambda, \Omega$ and $\Psi$:
\begin{align}
    \Lambda &= \frac{1}{K} \sum_{k=1}^K U^{(k)},\nonumber\\
    \Omega  &= \frac{1}{KD} \sum_{k = 1}^K (U^{(k)} - \Lambda)\Psi^{-1}(U^{(k)} - \Lambda)^\prime,\nonumber\\
    \Psi    &= \frac{1}{Kn} \sum_{k = 1}^K (U^{(k)} - \Lambda)^\prime \Omega^{-1}(U^{(k)} - \Lambda).
\end{align}
The estimates of $\Omega$ and $\Psi$ are computed in an iterative fashion until convergence. The likelihood of the matrix normal distribution remains the same if we scale $\Omega$ by a factor $c$ and multiple $\Psi$ by $1/c$. To circumvent this identifiability issue, we fix $\Omega_{11}$ to be 1 and solve for the maximum likelihood estimate (MLE) using a Lagrange multiplier \citep{glanz2018}. Under this constraint, the update of $\Omega$ consists of three steps. First, we calculate $S = \sum_{k = 1}^K (U_k - \Lambda) \Psi^{-1}(U_k - \Lambda)^\prime$. Second, we divide every entry in $S$ by $S_{11}$. Finally, we update the sub matrix $S_{-1,-1} = \frac{S_{11}}{KD} S_{-1, -1} + \left(1 - \frac{S_{11}}{KD}\right) S_{-1, 1} (S_{-1, 1})^\prime$, where $S_{-1,-1}$ is the matrix $S$ without the first row and column and $S_{-1, 1}$ is the first column vector of $S$ without the first element. We set the resulting matrix $S$ as the updated value of $\Omega$ in the current iteration.

Given the approximated posterior of $U|A,X$, we derive the marginal likelihood of the homophily-adjusted network effects model as follows.
\begin{align} 
    \nonumber
    y|A, X &\sim {{\mathcal N}} \Big((I - \rho A)^{-1}(\Lambda\gamma + X\beta), 
    \\
    &\qquad \qquad (I - \rho A)^{-1} \left( \gamma^\prime\Psi\gamma \Omega + \sigma^2 I \right)(I - \rho A')^{-1}\Big) \label{eq:marginal_effects}
\end{align}
Similarly, the marginal likelihood in the homophily-adjusted network disturbances model is given by
\begin{align}
    y|A, X &\sim {{\mathcal N}} \Big(\Lambda\gamma + X\beta,  \gamma^\prime \Psi \gamma \Omega + \sigma^2 (I_n - \rho A)^{-1}(I_n - \rho A')^{-1}\Big). 
    \label{eq:marginal_disturbances}
\end{align}
\noindent We will refer to the models given in (\ref{eq:marginal_effects}) and (\ref{eq:marginal_disturbances}) as the homophily-adjusted network effects (HANE) model and homophily-adjusted network disturbances (HAND) model respectively.

\subsection{Estimation}
\label{subsec:estimation}
For both the homophily-adjusted network effects and disturbances models, we obtain model estimates via a normal approximation to the posterior of  $(\beta, \gamma, \sigma^2, \rho)$ \citep{chen1985asymptotic}. In particular, we employed the limited memory BFGS algorithm with a box constraint to obtain to find the maximum a posterior (MAP) estimates and the Hessian \citep{byrd1995}. The prior distributions used in our analyses were
\begin{align}\nonumber
    \beta &\sim {{\mathcal N}} (0, \sigma^2_\beta I),
    \\ \nonumber
    \gamma &\sim {{\mathcal N}} (0, \sigma^2_\gamma I),
    \\ \nonumber
    \rho &\sim tr{\mathcal N}(\mu_\rho, \sigma_\rho{^2}, -1, 1),
    \\
    \sigma^2 &\sim IG\left(\frac{a}{2}, \frac{b}{2}\right),
\end{align}
where $tr{\mathcal N}(\mu, \sigma^2, a, b)$ is a truncated normal distribution with mean $\mu$, variance $\sigma^2$, and values restricted within $[a,b]$ range; $IG(c, d)$ is an inverse gamma distribution with shape $c$ and scale $d$. 

Details of log posterior and its gradients can be found in the Appendix. We initialized our estimation algorithm for the HANE and HAND models by using $\Lambda$ as a plug-in estimate of $U$ in Eqs. (\ref{eq:hane}) and (\ref{eq:disturbance_cond}) respectively and applying the two-stage least squares estimator proposed by \cite{kelejian1998}.
    
\section{Simulation study}
\label{sec:simstudy}

We conducted a simulation study to evaluate the performance of our proposed method. Simulating data involved two main steps. The first step was to generate a network. In order to emulate realistic networks, we first found an exemplar network, fit a generative model to it, and used this model output to then create new networks for our simulations.  The second step was to use the output from the first step to then generate new outcome variables.  The details of these two steps are given below.

As an exemplar network, we selected a real directed friendship network of 159 adolescence from the National Longitudinal Study of Adolescent to Adult Health (Add Health) study \citep{harris2011}. This network came from asking each student to list their closest five male and female friends. The data are publicly available as a part of the \textit{networkdata} package on GitHub \citep{networkdata}. We fit the latent cluster model \citep{handcock2007,krivitsky2009representing} to this dataset using the \textit{latentnet} package in R \citep{krivitsky2008, latentnet}, incorporating both random sender and receiver effects and setting the number of latent dimension $D=3$.  Using the cross-validation method proposed by \cite{Li2020}, we estimated the number of clusters in this network to be $K = 6$. We then kept the minimum Kullback-Leibler (mKL) estimates of the latent locations from the latent cluster model to assist in generating realistic networks.

We simulated new networks using both the mKL estimates of $U$ as described above and an additional node-level covariate, $X$, drawn from a normal distribution with a mean of 2 and a variance of 1. We used the absolute difference between the nodal attribute, that is, $|X_i-X_j|$, in generating the networks and assumed the coefficient of this term to be 0.5.

Given each simulated network $A$, $U$ from the mKL estimate, and $X$, we generated two outcome vectors corresponding to the limiting distributions given in Eqs. (\ref{eq:hane}) and (\ref{eq:disturbance_cond}). Note that the matrix normal approximation used in the HANE and HAND models we used to fit the simulated data was not a part of the simulation process. We assumed $\sigma^2 = 1$ and the intercept $\beta_0 = 0.5$. We varied the influence effect $\rho \in \{0, 0.1,\ldots, 0.6\}$, as well as the coefficients of $X$ $(\beta \in \{0, 0.5, 1\})$ and of $U$ $\left(\gamma^\prime \in \{(0.03, 0.05, -0.1), (0.06, 0.1, -0.2)\}\right)$. For each of these 84 scenarios ( $=$ 2(HANE/HAND) $\times7$($\rho$) $\times3$($\beta$) $\times2$($\gamma$)), we simulated 200 datasets for a total of 16,800 simulated datasets. 

Before applying our proposed method to each simulated dataset, we fit the latent cluster model to the simulated network, again using cross-validation to select the number of clusters, to obtain posterior samples of $U$ given the newly generated network and $X$. Note that due to the computational expense of fitting the latent cluster model to the data, we did not create a new network for every simulated dataset. Instead, we generated only 200 networks in the first step, each of which was fit using the latent cluster model. For each of the 84 scenarios, we used the same 200 simulated networks to then generate the 200 outcomes $y$ specific to that scenario. After simulating the data, we then fit the HANE or HAND model to the corresponding generated datasets, using the following hyperparameters: $\sigma_\beta = \sigma_\gamma = 2.25, \mu_\rho = 0.36, \sigma_\rho = 0.7, a = 2$ and $b = 2$. The prior on $\rho$ originated from \cite{dittrich2017bayesian} and was recommended in an extensive simulation study comparing different Bayesian and frequentist estimators of social influence by \cite{li2021}. The remaining hyperparemeters were chosen to induce weakly regularizing priors on $\beta, \gamma$ and $\sigma^2$. 

We compared our approach with both the Bayesian estimates and Maximum Likelihood Estimates (MLE) of the network autocorrelation model. Similar priors and estimation techniques using normal approximation were employed to derive the Bayesian estimates of NAM. To obtain the MLE of NAM, we used the \textit{sna} package in R \citep{butts2020}. The comparison metrics included bias, mean squared error (MSE), and the coverage of 95\% credible intervals for the influence measure $\rho$ and the coefficient of observed features $\beta$.

\begin{figure}[p]
    \centering
    \includegraphics[width = \textwidth]{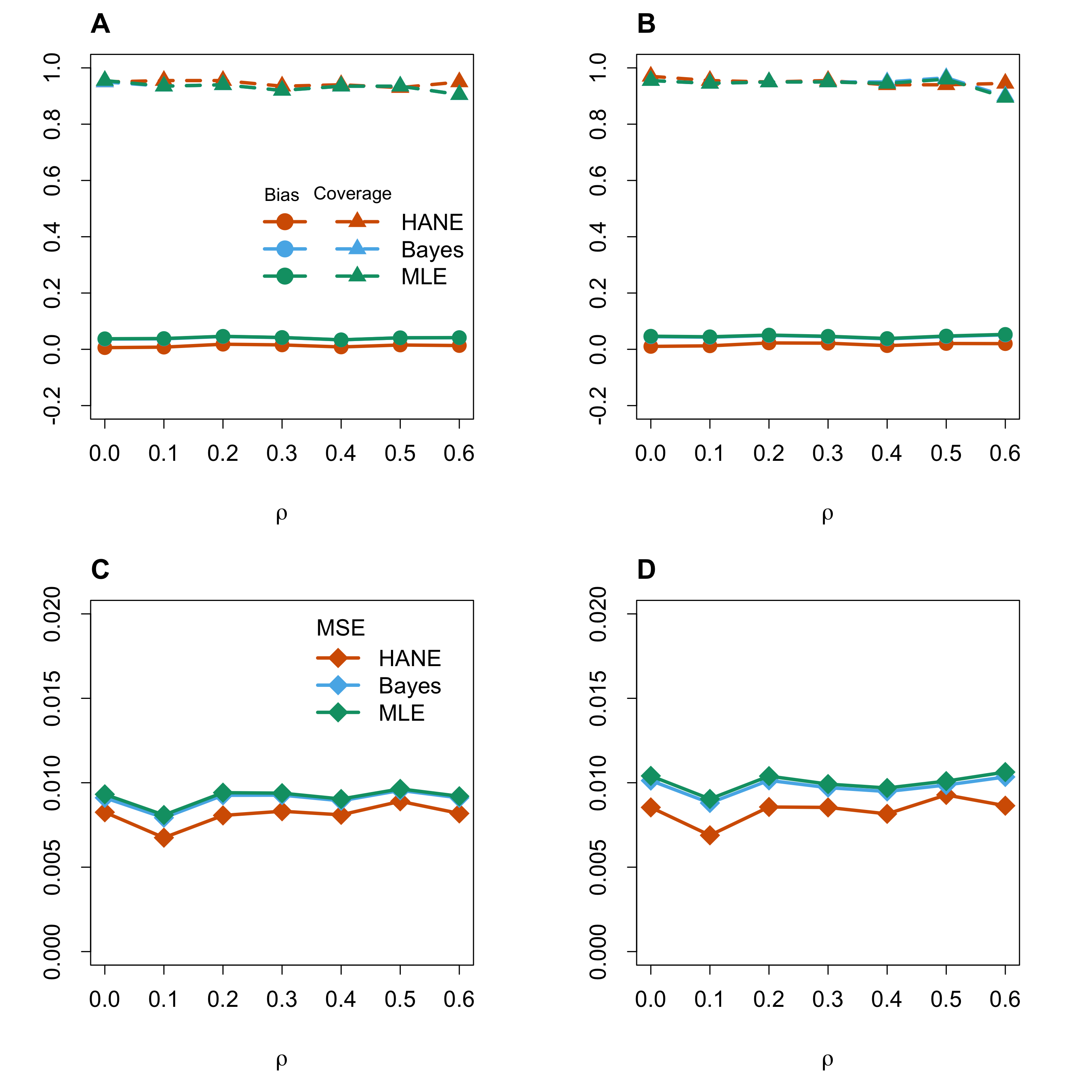}
%    \begin{subfigure}{0.45\textwidth}
%        \includegraphics[width=\textwidth]{hane_bias_b2_smallGamma.png}
%        \caption{Bias \& coverage given small $\gamma$}
%        \label{fig:beta_effect:a}
%    \end{subfigure}
%    \hfill
%    \begin{subfigure}{0.45\textwidth}
%        \includegraphics[width=\textwidth]{hane_bias_b2_largeGamma.png}
%        \caption{Bias \& coverage given large $\gamma$}
%        \label{fig:beta_effect:b}
%    \end{subfigure} \\
%    \begin{subfigure}{0.45\textwidth}
%        \includegraphics[width=\textwidth]{hane_MSE_b2_smallGamma.png}
%        \caption{MSE given small $\gamma$}
%        \label{fig:beta_effect:c}
%    \end{subfigure}
%    \hfill
%    \begin{subfigure}{0.45\textwidth}
%        \includegraphics[width=\textwidth]{hane_MSE_b2_largeGamma.png}
%        \caption{MSE given large $\gamma$}
%        \label{fig:beta_effect:d}
%    \end{subfigure}
    \caption{Comparing bias and 95\% confidence/credible interval coverage (A and B), and MSE (C and D) of $\beta$ estimates in the effects models, given $\beta=0.5$ while varying $\rho$ and $\gamma$. Columns correspond to the importance of the homophilic features ($\gamma$), where the first column (A and C) represent small $\gamma$ and the second column (B and D) reprent large $\gamma$. Color corresponds to different methods. Circles indicate the average bias, triangles indicates average coverage, and diamonds indicate MSE.}
    \label{fig:beta_effect}
\end{figure}

Due to the large number of scenarios, we have chosen to display the results only for $\beta=0.5$, as other scenarios yield similar plots. Those not included here are provided in the Supplementary Material.  

Figure \ref{fig:beta_effect} shows the bias, coverage rates, and MSE of the regression coefficient $\beta$ when varying $\rho$ and $\gamma$ in the effects models. Overall, the Bayesian estimates are similar to the MLEs in the NAM. The $\beta$ estimates from the HANE model had similar coverage rates but slightly smaller bias and MSE when compared to both estimates of the NAM. 

\begin{figure}[p]
    \centering
    \includegraphics[width = \textwidth]{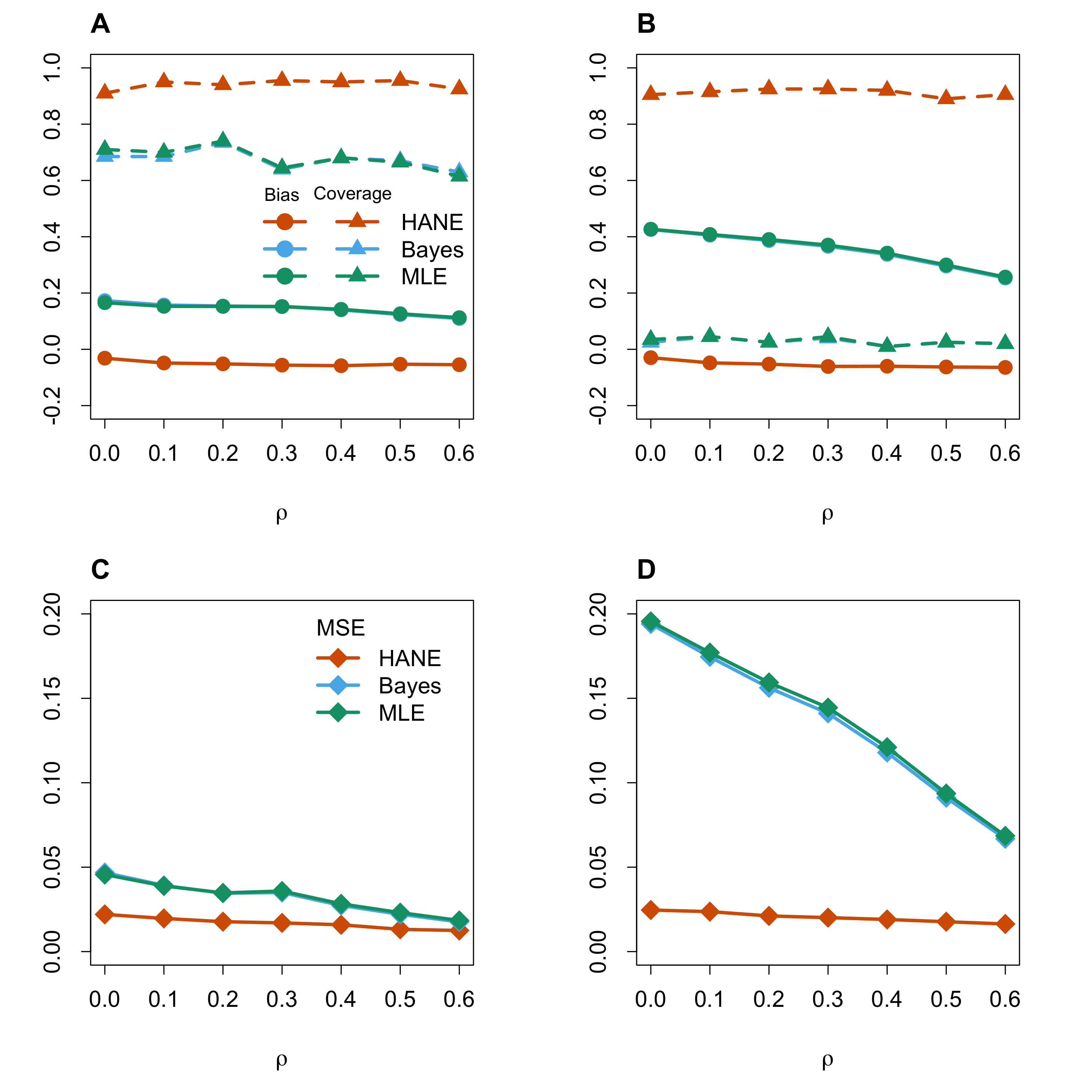}
%    \begin{subfigure}{0.45\textwidth}
%        \includegraphics[width=\textwidth]{hane_bias_rho_smallGamma.png}
%        \caption{Bias \& coverage given small $\gamma$}
%        \label{fig:rho_effect:a}
%    \end{subfigure}
%    \hfill
%    \begin{subfigure}{0.45\textwidth}
%        \includegraphics[width=\textwidth]{hane_bias_rho_largeGamma.png}
%        \caption{Bias \& coverage given large $\gamma$}
%        \label{fig:rho_effect:b}
%    \end{subfigure} \\
%    \begin{subfigure}{0.45\textwidth}
%        \includegraphics[width=\textwidth]{hane_MSE_rho_smallGamma.png}
%        \caption{MSE given small $\gamma$}
%        \label{fig:rho_effect:c}
%    \end{subfigure}
%    \hfill
%    \begin{subfigure}{0.45\textwidth}
%        \includegraphics[width=\textwidth]{hane_MSE_rho_largeGamma.png}
%        \caption{MSE given large $\gamma$}
%        \label{fig:rho_effect:d}
%    \end{subfigure}
    \caption{Comparing bias and 95\% confidence/credible interval coverage (A and B), and MSE (C and D) of $\rho$ estimates in the effects models, given $\beta=0.5$ while varying $\rho$ and $\gamma$. Columns correspond to the importance of the homophilic features ($\gamma$), where the first column (A and C) represent small $\gamma$ and the second column (B and D) reprent large $\gamma$. Color corresponds to different methods. Circles indicate the average bias, triangles indicates average coverage, and diamonds indicate MSE.}
    \label{fig:rho_effect}
\end{figure}

Figure \ref{fig:rho_effect} displays the bias, coverage rates, and MSE of the social influence parameter $\rho$ estimated when varying $\rho$ and $\gamma$ in the effects models. The HANE estimates of $\rho$ consistently outperformed both the Bayesian estimates and MLEs of the NAM, demonstrating smaller absolute bias, lower MSE, and improved coverage rates. Both the Bayesian estimates and MLEs of $\rho$ from the NAM suffered from a positive bias, which grew as $\rho$ decreased and $\gamma$ increased (see Figures \ref{fig:rho_effect:a}, \ref{fig:rho_effect:b}). In contrast, the $\rho$ estimate in the HANE model exhibited a slight negative bias that remained largely stable with any changes in $\rho$ and $\gamma$. 

\begin{figure}[p]
    \centering
    \includegraphics[width = \textwidth]{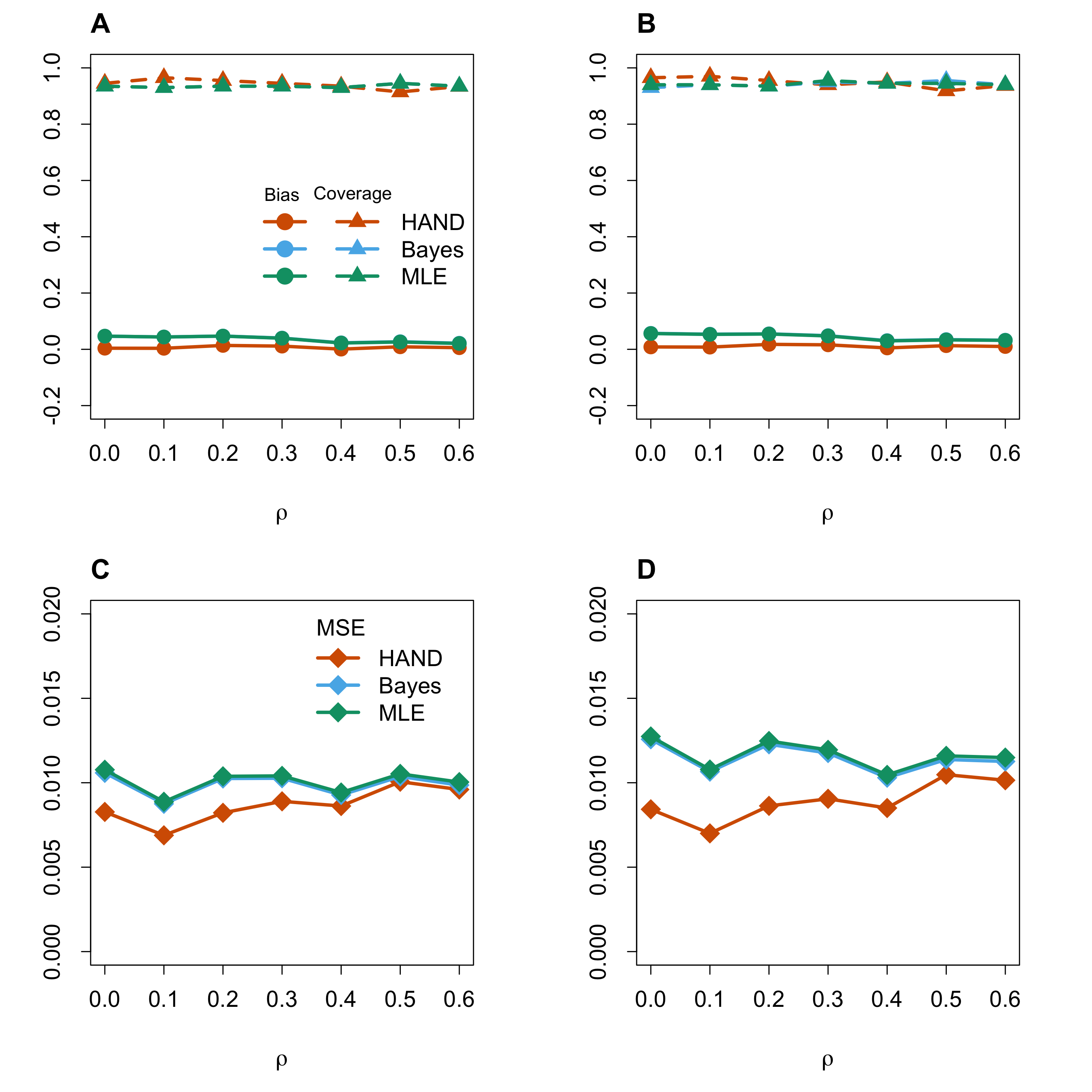}
%    \begin{subfigure}{0.45\textwidth}
%        \includegraphics[width=\textwidth]{hand_bias_b2_smallGamma.png}
%        \caption{Bias \& coverage given small $\gamma$}
%        \label{fig:beta_disturbance:a}
%    \end{subfigure}
%    \hfill
%    \begin{subfigure}{0.45\textwidth}
%        \includegraphics[width=\textwidth]{hand_bias_b2_largeGamma.png}
%        \caption{Bias \& coverage given large $\gamma$}
%        \label{fig:beta_disturbance:b}
%    \end{subfigure} \\
%    \begin{subfigure}{0.45\textwidth}
%        \includegraphics[width=\textwidth]{hand_MSE_b2_smallGamma.png}
%        \caption{MSE given small $\gamma$}
%        \label{fig:beta_disturbance:c}
%    \end{subfigure}
%    \hfill
%    \begin{subfigure}{0.45\textwidth}
%        \includegraphics[width=\textwidth]{hand_MSE_b2_largeGamma.png}
%        \caption{MSE given large $\gamma$}
%        \label{fig:beta_disturbance:d}
%    \end{subfigure}
    \caption{Comparing bias and 95\% confidence/credible interval coverage (A and B), and MSE (C and D) of $\beta$ estimates in the disturbances models, given $\beta=0.5$ while varying $\rho$ and $\gamma$. Columns correspond to the importance of the homophilic features ($\gamma$), where the first column (A and C) represent small $\gamma$ and the second column (B and D) reprent large $\gamma$. Color corresponds to different methods. Circles indicate the average bias, triangles indicates average coverage, and diamonds indicate MSE.}
    \label{fig:beta_disturbance}
\end{figure}

\begin{figure}[p]
    \centering
    \includegraphics[width = \textwidth]{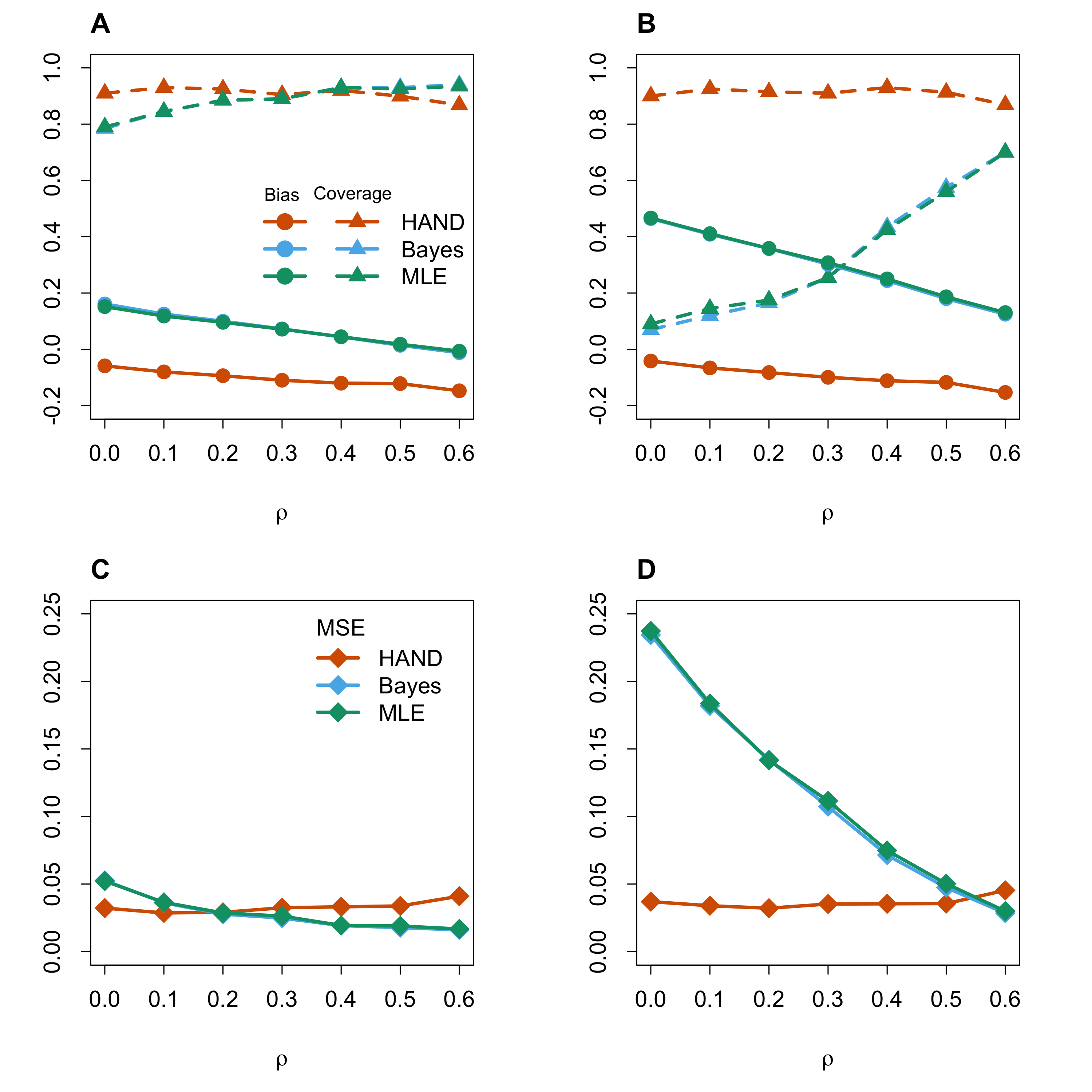}
%    \begin{subfigure}{0.45\textwidth}
%        \includegraphics[width=\textwidth]{hand_bias_rho_smallGamma.png}
%        \caption{Bias \& coverage given small $\gamma$}
%        \label{fig:rho_disturbance:a}
%    \end{subfigure}
%    \hfill
%    \begin{subfigure}{0.45\textwidth}
%        \includegraphics[width=\textwidth]{hand_bias_rho_largeGamma.png}
%        \caption{Bias \& coverage given large $\gamma$}
%        \label{fig:rho_disturbance:b}
%    \end{subfigure} \\
%    \begin{subfigure}{0.45\textwidth}
%        \includegraphics[width=\textwidth]{hand_MSE_rho_smallGamma.png}
%        \caption{MSE given small $\gamma$}
%        \label{fig:rho_disturbance:c}
%    \end{subfigure}
%    \hfill
%    \begin{subfigure}{0.45\textwidth}
%        \includegraphics[width=\textwidth]{hand_MSE_rho_largeGamma.png}
%        \caption{MSE given large $\gamma$}
%        \label{fig:rho_disturbance:d}
%    \end{subfigure}
    \caption{Comparing bias and 95\% confidence/credible interval coverage (A and B), and MSE (C and D) of $\rho$ estimates in the disturbances models, given $\beta=0.5$ while varying $\rho$ and $\gamma$. Columns correspond to the importance of the homophilic features ($\gamma$), where the first column (A and C) represent small $\gamma$ and the second column (B and D) reprent large $\gamma$. Color corresponds to different methods. Circles indicate the average bias, triangles indicates average coverage, and diamonds indicate MSE.}
    \label{fig:rho_disturbance}
\end{figure}

Similarly, in the disturbances models, our proposed HAND model produced $\beta$ estimates with marginally smaller bias and MSE compared to the NAM (see Figure \ref{fig:beta_disturbance}). Figure \ref{fig:rho_disturbance} illustrates the bias, coverage rates, and MSE of the social influence parameter $\rho$ estimate with varying $\rho$ and $\gamma$ in the disturbances models. The proposed HAND model provided superior $\rho$ estimates only for smaller values of true $\rho$, and its performance, while not as strong for higher values of true $\rho$, remained close to that of the NAM. Consistent with the effects models, the Bayesian estimates and MLEs of $\rho$ from NAM showed a persistent upward bias, which grew when $\rho$ decreased and $\gamma$ increased. On the other hand, estimates of $\rho$ from the HAND model displayed a downward bias unaffected by varying $\gamma$, yet this bias became more negative as $\rho$ increased. Importantly, the coverage rate of $\rho$ for the Bayesian and MLE interval estimates was only close to the nominal rate for very large $\rho$ and small $\gamma$, while this rate dipped as low as 12\% for large $\gamma$ and small $\rho$.  The HAND interval estimates of $\rho$ stayed near the nominal rate for all $\rho$ and $\gamma$.

\section{Application to real data}
\label{sec:application}

Physical activity plays a vital role in adult health by reducing the risk of chronic diseases, improving mental health, and enhancing overall quality of life \citep{Paluska2000, Warburton2006}. In this section, we illustrated our proposed method by applying it to a data set of friends and family to explore peer influence effect on activity levels. The original study tracked 130 adult residents of a young-family community living next to a research university in North America for over 15 months \citep{Aharony2011}. Data was collected through a combination of surveys and mobile-based signals, including accelerometry and communication activities \citep{Aharony2011}. To illustrate our method for cross-sectional data analysis, we utilized a combination of data aggregated over the course of the study and the baseline data where applicable. 

Using call log data, we constructed a directed network where an edge exists from node $i$ to $j$ if individual $i$ made a call to individual $j$ during the study period. The outcome of interest is the level of physical activity, which we determined using the average daily value collected from the FunFit app. Higher values indicate greater physical activity levels. Additionally, we included three baseline covariates in the social influence models: gender (male vs. female), sleep quality (5 hours or less, 6, 7, 8 hours, or more), and quality of life.  Quality of life (QoL) was calculated by averaging responses to eight Likert-scale items measuring life satisfaction (1=Strongly Disagree, 7=Strongly Agree). Before analysis, we standardized all continuous variables, including the physical activity outcome and the quality of life measure. Our analysis retained 120 community members with complete network and survey data.

Similar to the simulation study, we obtained posterior samples of the latent positions from the latent cluster LSM, using the same cross-validation approach to select number of clusters $(K = 3)$. All covariates included in the social influence models were also incorporated into the latent cluster LSM. Specifically, our model accounted for the number of edges where the incident nodes share the same gender and sleep quality, as well as the sum of the absolute differences in quality of life between the incident nodes for each edge. We used the same weakly regularizing priors as in the simulation study. We compared the results of our approach to Bayesian estimates and MLEs of NAMs without homophily adjustments. Table \ref{table:sim-results} shows the results of our proposed method alongside the others. While the estimates of the coefficients of observed features $\beta$ were similar between methods, the estimate of the social influence $\rho$ was substantially smaller in our approach, suggesting that those approaches that failed to adjust for the backdoor pathway of homophilic features overestimated the social influence at play.

\section{Discussion}
\label{sec:discussion}
The intertwining of homophily and social influence often leads to difficulty in distinguishing the two phenomenon. The current method of estimating social influence from cross-sectional data is the network autocorrelation model, whose justification has rested on the linear diffusion model. The assumptions of this deterministic model are too stringent for plausibility. In this paper, we have relaxed many of these assumptions, providing a corresponding class of homophily-adjusted network autocorrelation models for cross-sectional data. Within the Bayesian framework, we have derived parsimonious models for estimating social influence that leverage and are compatible with the wide array of latent space network models. We demonstrated via our simulation study that our methods resulted in influence estimates with smaller MSE and better credible interval coverage than methods which ignore additional homophilic features.

Despite the theoretical derivation and promising results, our models suffer from several limitations. Similar to other methods designed for cross-sectional data, our approach relies on strong assumptions regarding the stability of the network and individual behaviors over time. Further research is required to explore alternative methods that can relax these assumptions and to evaluate how the model's performance is affected when these assumptions are not met. Additionally, the normal approximation to the posterior may not be suitable for certain datasets due to potential boundary issues concerning $\rho$. Identifying a similarly fast yet more flexible Bayesian estimation method that can accommodate a broader range of datasets remains an area for future research.

%%%%%%%%%%%%%%

\begin{table}[htb]
	\begin{adjustbox}{width=\columnwidth,center}
		\begin{tabular}{lcccc}
			\hline
			\textbf{Model}                & \textbf{Variable}           & \textbf{Homophily Adjusted} & \textbf{Bayesian}   & \textbf{MLE}        \\ \hline
			\multirow{7}{*}{Effects}      & Intercept                   & 0.21 (-0.38, 0.79)          & 0.19 (-0.41, 0.79)  & 0.21 (-0.41, 0.83)  \\
			& $\beta_{\text{male}}$       & -0.08 (-0.38, 0.22)         & -0.08 (-0.40, 0.24) & -0.09 (-0.41, 0.24) \\
			& $\beta_{\text{sleep 6hr}}$  & 0.08 (-0.54, 0.69)          & -0.10 (-0.74, 0.54) & -0.12 (-0.78, 0.55) \\
			& $\beta_{\text{sleep 7hr}}$  & -0.21 (-0.80, 0.38)         & -0.32 (-0.94, 0.30) & -0.34 (-0.98, 0.30) \\
			& $\beta_{\text{sleep 8+hr}}$ & -0.34 (-0.96, 0.28)         & -0.44 (-1.08, 0.21) & -0.46 (-1.13, 0.21) \\
			& $\beta_{\text{QoL}}$        & -0.08 (-0.24, 0.08)         & -0.10 (-0.26, 0.06) & -0.10 (-0.26, 0.06) \\
			& $\rho$                      & 0.32 (0.10, 0.55)           & 0.43 (0.22, 0.63)   & 0.43 (0.22, 0.63)   \\ 
%			\hline
			\cline{2-5}
			\multirow{7}{*}{Disturbances} & Intercept                   & 0.15 (-0.48, 0.78)          & -0.03 (-0.72, 0.65) & -0.03 (-0.74, 0.68) \\
			& $\beta_{\text{male}}$       & -0.12 (-0.41, 0.17)         & -0.20 (-0.50, 0.10) & -0.20 (-0.50, 0.11) \\
			& $\beta_{\text{sleep 6hr}}$  & 0.11 (-0.48, 0.70)          & -0.01 (-0.62, 0.59) & -0.02 (-0.64, 0.61) \\
			& $\beta_{\text{sleep 7hr}}$  & -0.11 (-0.70, 0.48)         & -0.09 (-0.71, 0.53) & -0.10 (-0.74, 0.55) \\
			& $\beta_{\text{sleep 8+hr}}$ & -0.29 (-0.89, 0.31)         & -0.36 (-0.98, 0.26) & -0.37 (-1.01, 0.27) \\
			& $\beta_{\text{QoL}}$        & -0.10 (-0.26, 0.06)         & -0.11 (-0.27, 0.06) & -0.11 (-0.27, 0.06) \\
			& $\rho$                      & 0.34 (0.10, 0.57)           & 0.47 (0.26, 0.68)   & 0.47 (0.26, 0.68)   \\ \hline
		\end{tabular}
	\end{adjustbox}
	\caption{Estimates and 95\% credible intervals of coefficients $\beta$ and social influence $\rho$ in the real data application.}
	\label{table:sim-results}
\end{table}

%%%%%%%%%%%%%%

\begin{appendices}

\section{A1. Proof of Theorem 3.1} %\ref{theorem:effect}}
\label{appendix:hane_theorem}
We expand on the outcomes at previous times in Eq. (6) 
%(\ref{eq:hane_long}) 
to obtain:
\begin{align*}
y_t &=U\gamma + X\beta + \rho A y_{t-1} + \sigma_{\alpha,t}\alpha + \sigma_{\epsilon,t}\epsilon_t \\
&= \sum_{j = 0}^t \rho^j A^j (U\gamma + X\beta) + p^t A^t y_0 + \sum_{j = 0}^t \rho^j A^j \sigma_{\alpha, t - j} \alpha + \sum_{j = 0}^t \rho^j A^j \sigma_{\epsilon, t - j}\epsilon_{t-j}\\
&= \sum_{j = 0}^t \rho^j A^j (U\gamma + X\beta) + \sum_{j = 0}^t \rho^j A^j \sigma_{\alpha}\alpha + p^t A^t y_0 + \sum_{j = 0}^t \rho^j A^j (\sigma_{\alpha, t - j} - \sigma_\alpha)\alpha\\
&\quad+ \sum_{j = 0}^t \rho^j A^j \sigma_{\epsilon, t - j}\epsilon_{t-j}
\end{align*}

It follows from Assumption 1
%\ref{assump:convergence} 
that $\sum_{j = 0}^\infty \rho^j A^j \to (I - \rho A)^{-1}$ as $t \to \infty$. Therefore, we conclude that $\sum_{j = 0}^t \rho^j A^j (U\gamma + X\beta) \overset{a.s.}{\to} (I - \rho A)^{-1}(U\gamma + X\beta)$ and $\sum_{j = 0}^t \rho^j A^j \sigma_{\alpha}\alpha \overset{a.s.}{\to} (I - \rho A)^{-1}\sigma_{\alpha}\alpha$. In addition, Assumption 1 
%\ref{assump:convergence} 
implies that $p^t A^t \to 0$ as $t \to \infty$. By Assumption 6 
%\ref{assump:finiteness}
, it follows that 
$p^t A^t y_0 \overset{a.s.}{\to} 0$.

We will now show that $\sum_{j = 0}^t \rho^j A^j (\sigma_{\alpha, t - j} - \sigma_\alpha)\alpha \overset{a.s.}{\to} 0$ by showing that $\sum_{j = 0}^t \rho^j A^j (\sigma_{\alpha, t - j} - \sigma_\alpha)$ converges point-wise to $0$. For the remainder of the proof, let inequalities denote element-wise inequalities. First, we show via induction that $A^\ell \leq J \forall \ell \geq 0$, where $J$ is the $n \times n$ matrix of 1. Let $(A^\ell)_{ik}$ denote the $[i,k]$ element of $A^\ell$. For the base cases, $(A^0)_{ik} = (I)_{ik} \leq 1$ and $(A^1)_{ik} = (A)_{ik} \leq 1 \forall \{i,k\}$, since $A$ is a row normalized adjacency matrix. Assume $(A^\ell)_{ik} \leq 1 \forall \{i,k\}$, we obtain
\begin{align*}
(A^{\ell+1})_{ik} = (AA^\ell)_{ik} = \sum_{h = 1}^n (A)_{ih}(A^\ell)_{hk} \leq \sum_{h = 1}^n (A)_{ih}\times1 = \sum_{h = 1}^n (A)_{ih} \leq 1.
\end{align*}
Therefore, $A^\ell \leq J \forall \ell \geq 0$. Similarly, we also have $(A^{'})^\ell \leq J \forall \ell \geq 0$.\\

Then, we prove the point-wise convergence. Let $m$ denote the bound of both $(\sigma_{\alpha,t} - \sigma_\alpha)$ and $\sigma_{\epsilon, t}$. Since $A$ is row-normalized, Assumption 1 
%\ref{assump:convergence} 
implies that $|\rho| < 1$. As a result, $\sum_{j \geq 0} |\rho|^j = \frac{1}{1 - |\rho|}$ (by geometric series) and for any $\delta > 0$, $\exists t_\rho$ such that $|\rho|^j < \frac{\delta (1 - |\rho|)}{2m}  \forall j \geq t_\rho$. Furthermore, by Assumption 7
%\ref{assump:stability_deviations}
, $\exists t_\alpha$ such that $|\sigma_{\alpha,t} - \sigma_\alpha| < \frac{\delta(1 - |\rho|)}{2} \forall t\geq t_\alpha$. Finally, let $T:= t_\alpha + t_\rho$.\\

Let $[A]^\text{abs}$ denote the element-wise absolute values of A. We obtain the following result.
\begin{align*}
\left[\sum_{j = 0}^T \rho^jA^j(\sigma_{\alpha,T-j} - \sigma_\alpha)\right]^\text{abs} &\leq \sum_{j = 0}^{t_\rho-1} |\rho|^jA^j|\sigma_{\alpha,T-j} - \sigma_\alpha| + \sum_{j = t_\rho}^{T} |\rho|^jA^j|\sigma_{\alpha,T-j} - \sigma_\alpha|\\
&< \frac{\delta(1 - |\rho|)}{2}\sum_{j = 0}^{t_\rho-1} |\rho|^j J + m\sum_{j = t_\rho}^{T} |\rho|^j J\\
&= \frac{\delta(1 - |\rho|)}{2}\left(\sum_{j = 0}^{t_\rho-1} |\rho|^j \right) J + m|\rho|^{t_\rho} \left( \sum_{j = 0}^{T-t_\rho} |\rho|^j\right)J\\
&< \frac{\delta(1 - |\rho|)}{2}\left(\sum_{j = 0}^{\infty} |\rho|^j\right) J + m\frac{\delta (1 - |\rho|)}{2m} \left(\sum_{j = 0}^{\infty} |\rho|^j\right) J\\
&=\frac{\delta}{2}J + \frac{\delta}{2}J\\
&= \delta J
\end{align*}

Therefore, $\sum_{j = 0}^t \rho^j A^j (\sigma_{\alpha, t - j} - \sigma_\alpha)$ converges element-wise to 0. We conclude that $\sum_{j = 0}^t \rho^j A^j (\sigma_{\alpha, t - j} - \sigma_\alpha)\alpha \overset{a.s.}{\to} 0$.\\

Finally, we show that $\sum_{j = 0}^t \rho^j A^j \sigma_{\epsilon, t - j}\epsilon_{t-j} \overset{{\mathcal D}}{\to} 0$ as $t \to \infty$. The idea is to show that the characteristic function of $S_t = \sum_{j = 0}^t \rho^j A^j \sigma_{\epsilon, t - j}\epsilon_{t-j}$ converges point-wise to $1$. In particular, we need to show that
\begin{align*}
\phi_{S_t}(z) &= 
\exp\left\{
-\frac12 z' \left[
\sum_{j=0}^t\rho^{2j}A^{j}(A')^{j}\sigma^2_{\epsilon,t-j}
\right] z
\right\} \to 1,
\end{align*}
which is equivalent to showing that $\sum_{j=0}^t\rho^{2j}A^{j}(A')^{j}\sigma^2_{\epsilon,t-j}$ converges point-wise to $0$. The proof of the latter is as follows.

As mentioned above, $|\rho| < 1$ so $\sum_{j \geq 0} (\rho^2)^j = \frac{1}{1 - \rho^2}$ (by geometric series). For any $\delta > 0$, $\exists t_\rho$ such that $(\rho^2)^j < \frac{\delta (1 - \rho^2)}{2nm^2}  \forall j \geq t_\rho$. Furthermore, by Assumption 7
%\ref{assump:stability_deviations}
, $\exists t_\epsilon$ such that $\sigma_{\epsilon, t}^2 < \frac{\delta(1 - \rho^2)}{2n} \forall t\geq t_\epsilon$. Finally, let $T:= t_\epsilon + t_\rho$. Notice that
\begin{align*}
\sum_{j=0}^T\rho^{2j}A^{j}(A')^{j}\sigma^2_{\epsilon,T-j} &= \sum_{j = 0}^{t_\rho-1} \rho^{2j} A^{j}(A')^{j} \sigma^2_{\epsilon,T-j} + \sum_{j = t_\rho}^{T} \rho^{2j}A^{j}(A')^{j}\sigma^2_{\epsilon,T-j}\\
&< \frac{\delta(1 - \rho^2)}{2n}\sum_{j = 0}^{t_\rho-1} \rho^{2j} J J + m^2\sum_{j = t_\rho}^{T} \rho^{2j} JJ\\
&= \frac{\delta(1 - \rho^2)}{2}\left(\sum_{j = 0}^{t_\rho-1} (\rho^2)^j \right) J + m^2 (\rho^2)^{t_\rho} \left(\sum_{j = 0}^{T - t_\rho} (\rho^2)^j \right) JJ\\
&< \frac{\delta(1 - \rho^2)}{2}\left(\sum_{j = 0}^{\infty} (\rho^2)^j \right) J + m^2\frac{\delta (1 - \rho^2)}{2nm^2} \left(\sum_{j = 0}^{\infty} (\rho^2)^j\right) nJ\\
&=\frac{\delta}{2}J + \frac{\delta}{2}J\\
&= \delta J.
\end{align*}

We conclude that characteristic function $\phi_{S_t}(z)$ converges point-wise to 1. It follows by L\'evy's convergence theorem that $S_t$ converges in distribution to $0$. \\

Since almost sure convergence implies convergence in probability, it follows by Slutsky's theorem that $y_t\overset{{\mathcal D}}{\to} N\left( (I-\rho A)^{-1}(U\gamma + X\beta), \sigma^2_\alpha (I-\rho A)^{-1}(I-\rho A')^{-1}\right)$.

\section{A2. Gradients of the log posterior of the homophily adjusted network effect and disturbance models}
\label{appendix:gradients}
In the HANE model, denote $M = (I - \rho A)^{-1}$ and the variance covariance $\Sigma_e = \gamma^\prime \Psi \gamma \Omega + \sigma^2 I$. The log posterior up to a constant is given by
\begin{align*}
l_e &\propto  -\frac{1}{2} \log \det{\Sigma_e} - \log \det{M} - \frac{(M^{-1}y - \Lambda\gamma - X\beta)^\prime\Sigma_e^{-1}(M^{-1}y - \Lambda\gamma - X\beta)}{2} \nonumber \\
&\qquad - \frac{\beta^\prime\beta}{2\sigma_\beta^2} - \frac{\gamma^\prime\gamma}{2\sigma_\gamma^2} - \frac{(\rho - \mu_{\rho})^2}{2\sigma_{\rho}^2}+ (-\frac{a}{2}-1)\log(\sigma^2) -\frac{b}{2\sigma^2}.
\end{align*}
The gradients of the log posterior of the HANE model are given by
\begin{align*}
\frac{\delta l_e}{\delta \beta} &= X^\prime \Sigma_e^{-1}(M^{-1}y - \Lambda \gamma - X \beta) - \sigma^{-2}_\beta \beta ,\\
\frac{\delta l_e}{\delta \gamma_i} &=  \Lambda_j^\prime \Sigma_e^{-1}(M^{-1}y - \Lambda\gamma - X \beta) - \sigma^{-2}_\gamma \gamma_j - \frac{1}{2} Tr(2[\Psi\gamma]_j\Omega)  \nonumber\\
&\quad + \frac{(M^{-1}y - \Lambda\gamma - X\beta)^\prime\Sigma_e^{-1}2[\Psi\gamma]_j\Omega\Sigma_e^{-1}(M^{-1}y - \Lambda\gamma - X\beta)}{2}, \\
\frac{\delta l_e}{\delta \rho} &= -Tr(AM) + (Ay)^\prime\Sigma_e^{-1}(M^{-1}y - \Lambda\gamma - X\beta) -\frac{\rho - \mu_{\rho}}{\sigma_{\rho}^2},\\
\frac{\delta l_e}{\delta \sigma^2} &= -\frac{1}{2} Tr(\Sigma_e^{-1}) + \frac{(M^{-1}y - \Lambda\gamma - X\beta)^\prime\Sigma_e^{-1}\Sigma_e^{-1}(M^{-1}y - \Lambda\gamma - X\beta)}{2} \\
&\quad - \left(\frac{a}{2} + 1 \right)\frac{1}{\sigma^2} + \frac{b}{2(\sigma^2)^2},
\end{align*}
where $\Lambda_j$ denotes the $j^{th}$ column of $\Lambda$, $\gamma_j$ denotes the $j^{th}$ element of vector $\gamma$ and $[\Psi\gamma]_j$ denotes the $j^{th}$ element of vector $\Psi\gamma$. 

In the HAND model, denote the variance covariance $\Sigma_d = \gamma^\prime \Psi \gamma \Omega + \sigma^2 (I - \rho A)^{-1}(I - \rho A^\prime)^{-1}$. The log posterior up to a constant is given by
\begin{align*}
l_d &\propto  -\frac{1}{2} \log \det{\Sigma_d} - \frac{(y - \Lambda\gamma - X\beta)^\prime\Sigma_d^{-1}(y - \Lambda\gamma - X\beta)}{2} \nonumber\\
&\qquad - \frac{\beta^\prime\beta}{2\sigma_\beta^2} - \frac{\gamma^\prime\gamma}{2\sigma_\gamma^2} -\frac{(\rho - \mu_{\rho})^2}{2\sigma_{\rho}^2} + (-\frac{a}{2}-1)\log(\sigma^2) -\frac{b}{2\sigma^2}.
\end{align*}
The gradients of the log posterior of the HAND model are given by
\begin{align*}
\frac{\delta l_d}{\delta \beta} &= X^\prime\Sigma_d^{-1}(y - \Lambda \gamma - X \beta) - \sigma^{-2}_\beta \beta,\\
\frac{\delta l_d}{\delta \gamma_i} &=  \Lambda_i^\prime \Sigma_d^{-1}(y - \Lambda\gamma - X \beta) - \sigma^{-2}_\gamma \gamma_i -\frac{1}{2} Tr\left(2[\Psi\gamma]_i\Sigma_d^{-1}\Omega\right) \nonumber\\
&\quad + \frac{(y - \Lambda\gamma - X\beta)^\prime\Sigma_d^{-1}2[\Psi\gamma]_i\Omega\Sigma_d^{-1}(y - \Lambda\gamma - X\beta)}{2},\\
\frac{\delta \l_d}{\delta \rho} &= -\frac{1}{2} Tr\left(\sigma^2\Sigma_d^{-1}M(AM + M^\prime A^\prime)M^\prime\right) -\frac{\rho - \mu_{\rho}}{\sigma_{\rho}^2} \nonumber\\
&\quad + \frac{\sigma^2(y - \Lambda\gamma - X\beta)^\prime\Sigma_d^{-1}M(AM + M^\prime A^\prime)M^\prime\Sigma_d^{-1}(y - \Lambda\gamma - X\beta)}{2},\\
\frac{\delta l_d}{\delta \sigma^2} &= -\frac{1}{2} Tr\left(\Sigma_d^{-1}MM^\prime\right) - (\frac{a}{2} + 1 )\frac{1}{\sigma^2} + \frac{b}{2(\sigma^2)^2} \\
&\quad  + \frac{(y - \Lambda\gamma - X\beta)^\prime\Sigma_d^{-1}MM^\prime\Sigma_d^{-1}(y - \Lambda\gamma - X\beta)}{2}.
\end{align*}

\end{appendices}

%%%%%%%%%%%%%%

\section{Competing interests}
No competing interest is declared.

%%%%%%%%%%%%%%

%\section{Author contributions statement}
%
%Must include all authors, identified by initials, for example:
%S.R. and D.A. conceived the experiment(s),  S.R. conducted the experiment(s), S.R. and D.A. analysed the results.  S.R. and D.A. wrote and reviewed the manuscript.

%%%%%%%%%%%%%%

%\section{Acknowledgments}
%The authors thank the anonymous reviewers for their valuable suggestions. This work is supported in part by funds from the National Science Foundation (NSF: \# 1636933 and \# 1920920).

%%%%%%%%%%%%%%

\bibliographystyle{abbrvnat}
\bibliography{hnam}

\end{document}

\section*{Complete simulation results for ``Homophily-adjusted social influence estimation''}

\begin{figure}[htb!]
    \centering
    \includegraphics[width=\linewidth]{hane_all_bias_b2.png}
    \caption{Comparing bias and 95\% confidence/credible interval coverage of $\beta$ in the effects models. Color corresponds to different methods. Circles indicate the average bias and triangles indicates average coverage.}
\end{figure}

\begin{figure}[htb!]
    \centering
    \includegraphics[width=\linewidth]{hane_all_MSE_b2.png}
    \caption{Comparing MSE of $\beta$ in the effects models. Color corresponds to different methods. Circles indicate the average bias and triangles indicates average coverage.}
\end{figure}

\begin{figure}[htb!]
    \centering
    \includegraphics[width=\linewidth]{hane_all_bias_rho.png}
    \caption{Comparing bias and 95\% confidence/credible interval coverage of $\rho$ in the effects models. Color corresponds to different methods. Circles indicate the average bias and triangles indicates average coverage.}
\end{figure}

\begin{figure}[htb!]
    \centering
    \includegraphics[width=\linewidth]{hane_all_MSE_rho.png}
    \caption{Comparing MSE of $\rho$ in the effects models. Color corresponds to different methods. Circles indicate the average bias and triangles indicates average coverage.}
\end{figure}

\begin{figure}[htb!]
    \centering
    \includegraphics[width=\linewidth]{hand_all_bias_b2.png}
    \caption{Comparing bias and 95\% confidence/credible interval coverage of $\beta$ in the disturbances models. Color corresponds to different methods. Circles indicate the average bias and triangles indicates average coverage.}
\end{figure}

\begin{figure}[htb!]
    \centering
    \includegraphics[width=\linewidth]{hand_all_MSE_b2.png}
    \caption{Comparing MSE of $\beta$ in the disturbances models. Color corresponds to different methods. Circles indicate the average bias and triangles indicates average coverage.}
\end{figure}

\begin{figure}[htb!]
    \centering
    \includegraphics[width=\linewidth]{hand_all_bias_rho.png}
    \caption{Comparing bias and 95\% confidence/credible interval coverage of $\rho$ in the disturbances models. Color corresponds to different methods. Circles indicate the average bias and triangles indicates average coverage.}
\end{figure}

\begin{figure}[htb!]
    \centering
    \includegraphics[width=\linewidth]{hand_all_MSE_rho.png}
    \caption{Comparing MSE of $\rho$ in the disturbances models. Color corresponds to different methods. Circles indicate the average bias and triangles indicates average coverage.}
\end{figure}